\documentclass[preprint,showpacs,preprintnumbers,amsmath,amssymb]{revtex4}

\usepackage{graphicx}
\usepackage{dcolumn}
\usepackage{bm}

\begin{document}


\title{Influence of electron capture and Coulomb explosion 
on electron screening in low energy nuclear reactions in laboratories}

\author{Yasuo Kato}
 \email{kato@nucl.phys.tohoku.ac.jp}
\author{Noboru Takigawa}%
 \email{takigawa@nucl.phys.tohoku.ac.jp}
\affiliation{%
Department of Physics, Graduate School 
of Science, Tohoku University, Sendai 980-8578,Japan}%

\date{\today}

\begin{abstract}
We discuss the effects of electron capture by the projectile 
and the Coulomb explosion of a molecular projectile 
on the electron screening in low energy nuclear
reactions in laboratory. 
Using the idea of equilibrium charge, we 
show that the electron capture of projectile 
leads to a screening energy 
which significantly exceeds the adiabatic limit in the simple consideration 
for the D(d,p)T reaction 
and provides a possibility to explain the large screening 
energy claimed in the analysis of experimental data. 
We then show that the Coulomb explosion can result in a 
large apparent screening energy 
as large as that encountered 
in the analysis of  $^3${\rm He}(d,p)$^4${\rm He} reactions induced by the 
molecular $\rm D^+_2$ and $\rm D^+_3$ projectiles at very low energies.
\end{abstract}


\maketitle

\section{Introduction}

Nuclear reaction rates at the Gamow energy play a key role in  
the synthesis of elements and generation of energy in
stars. However, it is difficult to determine them directly 
by experiments in laboratories because of the small cross section 
due to the tunneling procedure through the Coulomb barrier. 
One then tries to determine them by extrapolating the reaction rates 
observed at high energies to lower energies. The extrapolation is 
usually done for the astrophysical $S(E)$ factor introduced by 
\begin{equation}
\sigma(E)=\frac{S(E)}{E}\exp(-2\pi\eta)
\label{s}
\end{equation}
with the Sommerfeld parameter $\eta$=Z$_1$Z$_2e^2/\hbar v$,
Z$_1$ and Z$_2$ being the atomic numbers of the projectile and 
target nuclei and $v$ the initial velocity of the collision, 
and assuming that $S(E)$ depends only weakly on energy and 
can be well parametrized with a low order polynomial 
unless resonance states 
are involved.
 
As the measurements are extended to lower energies, the 
observed values have been found to 
be significantly larger than those predicted by the 
extrapolation of the cross section at high energies
\cite{alr87}. The 
enhancement gets larger with decreasing collision energy.
Many works have been reported which try to attribute this phenomenon to 
the screening effect by bound electrons in the target and in some cases 
also in the projectile. Assuming that the 
screening effect can be well represented by a spatially 
constant lowering of the 
Coulomb barrier by the amount of $U_e$, 
one often expresses the enhancement of the cross section by
\begin{align}
f&\equiv\frac{\sigma(E)}{\sigma_0(E)}=\frac{\sigma_0(E+U_e)}
{\sigma_0(E)}\nonumber\\
&=\frac{S(E+U_e)}{S(E)}\frac{E}{E+U_e}\frac{\exp[-2\pi
\eta(E+U_e)]}{\exp[-2\pi\eta(E)]}\nonumber\\
&\simeq\exp\left\{\pi\eta(E)\frac{U_e}{E}\right\}.
\label{zou}
\end{align}
Here, $\sigma(E)$ and $\sigma_0(E)$ are the true cross section and 
the cross section in the absence of screening effects, respectively. 
It was assumed that $U_e\ll E$ and $S(E)$ is almost energy independent. 
The $U_e$ is called the screening energy. 

A puzzle is that the value of screening energy which was determined 
by fitting the experimental data with eq.(\ref{zou}) 
exceeds the theoretical value in the so called adiabatic limit, 
which is given by the 
difference of the binding energies of electrons 
in the united atom and in the initial state in the target or projectile 
and is thought to provide the maximum screening energy, for all
systems so far studied experimentally \cite{rs95}. 
In a recent paper \cite{ktab03} one of the authors of the present paper 
has discussed the influence of  
tunneling on electron screening and pointed out that the electron screening 
can exceed the adiabatic limit 
if the electronic state is not a
single adiabatic state at the external turning point either by pre-tunneling
transitions of the electronic state or by the symmetry of the system. 
However, the amount of excess is negligibly small to explain the large 
experimental value of the screening energy. 
Alternatively, the stopping power 
at low energies \cite{gs-e,gs-t,bp} and also 
the values of the screening energy \cite{barker,junker} have recently 
been reexamined.

In this paper we reexamine the effects of electron 
capture by the projectile and of the Coulomb explosion which seem to have been 
omitted in most recent works since the pioneering works 
\cite{bra90} and \cite{eng88}, respectively. In Sect. II, we collect a few basic formulae 
for the screening energy. 
In Sects. III and IV we discuss the effects of electron capture 
and the Coulomb explosion, respectively. 
We summarize the paper in Sect. V. 

\section{Screening energy in the presence of admixture of adiabatic states}

The Coulomb explosion leads to a spread in 
the energy of the nuclear reaction, while the electron capture 
to an admixture of electronic states. Denoting the screening energy in the 
n-th state and the corresponding mixing probability as $U_e^{(n)}$ and 
$P_n$, respectively, we evaluate the net enhancement factor by 
\begin{equation}
f=\frac{\sum_{n}P_n\times\sigma_0(E+U_e^{(n)})}{\sigma_0(E)}.
\label{true}
\end{equation}
It is then converted into the screening energy by 
\begin{equation}
U_e=\frac{E}{\pi\eta(E)}\log\sum_n\left[P_n\exp\left\{\pi\eta(E)
\frac{U_e^{(n)}}{E}\right\}\right].
\label{average}
\end{equation}
We note that the screening energy estimated by eq.(\ref{average}) is 
larger than a simpler estimate, where the enhancement factor is 
evaluated as 
\begin{equation}
f=\frac{\sigma_0(E+\sum_nP_n\times U_e^{(n)})}{\sigma_0(E)}.
\label{fcon}
\end{equation}
and consequently the screening energy is given by
\begin{equation}
U_e=\sum_nP_n U_e^{(n)}.
\label{secon}
\end{equation}
Eqs.(\ref{fcon}) and (\ref{secon}) ignore the variation of the 
tunneling probability in each channel \cite{ktab03}.

\section{Effects of electron capture}

We first discuss the effect of electron capture by the projectile. 
The charge state of the projectile can be different from the initial one 
at the time when it reacts with the target. 
This effect will become increasingly 
important as the energy gets lower. It is related to the fact that 
the stopping power of a charged particle passing through matter 
originates mainly from the exchange of 
electrons between the incoming charged particle and the surroundings 
in some low energy region \cite{bp,pdg}. 
One could thus use a similar technique in order 
to study the role of electron capture in the screening problem.  
For example, a simple estimate of the probability of the electron 
capture by the projectile and the charge exchange 
between the projectile and the 
target material could be obtained by solving the time evolution of 
the system consisting of the incident ion and a neutral target with a 
single valence electron. It suggests 
a large probability of electron capture by the projectile at low energies 
\cite{bp,kato04}.
In this paper, we resort to the idea of the equilibrium charge \cite{all58}
in order to estimate the average charge state of the projectile when it 
reacts with the target. 

Let us consider D(d,p)T reaction at $E=1.62\,\rm keV$, which is the 
lowest energy where experimental measurements were ever performed, 
as the first example. 
According to \cite{all58}, the incoming deuteron captures an electron 
to become a neutral deuterium atom D in the probability of 
about 90 $\%$ and remains to 
be deuteron in only 10 $\%$ probability at this energy. 
In the latter case, eq.(\ref {average}) 
gives the screening energy to be $U_e=22.0\,\rm eV$ 
by taking the admixture of the gerade and ungerade configurations 
with equal weight due to the symmetry of the system into account\cite{ktab03}. 
In order to evaluate the screening energy 
for the former, i.e. for the D+D reaction, we notice that 
the whole colliding system will 
converge to the ground and to the first excited states of He atom in the 
adiabatic limit with the probability of 
1/4 and 3/4, respectively, reflecting the 
statistical weights of the total spin of the two electrons. 
Referring to \cite{wei,com} for the binding energy of electrons, 
the screening energy in each charge state is given by, 
\begin{align*}
78.9-(13.6+13.6)&=51.7\,[\rm eV]\;\;\;(\mbox{ground state})\\
59.1-(13.6+13.6)&=31.9\,[\rm eV]\;\;\;(\mbox{first excited state}).
\end{align*}
Eq.(\ref {average}) then leads to the screening energy $U_e=37.1\,\rm eV$ 
for this channel at $E=1.62\,\rm keV$. 
The net screening energy, which is observed experimentally,
should be given by the average of these two charge states with the 
weights of 1:9. Using eq.(\ref{average}), we finally obtain 
$U^{(c)}_e$=35.7eV, where 
the upper suffix (c) stands for capture of electrons by the 
projectile. 

The second example of the effect of electron capture by the projectile 
is the $^3$He(d,p)$^4$He reaction at $E=5.01\,\rm keV$. In this reaction, 
about $10\sim20$ $\%$ of the projectile captures an electron. 
The D and d projectile lead to the ground states of Li and Li$^+$, 
respectively. The screening energy for each case is 
\begin{align*}
203.4-(78.9+13.6)&=110.9\,[\rm eV]\;\;\;(D+ {^3\rm{He}})\\
198.0-78.9\;\;\;\;\;\;\;&=119.1\,[\rm eV]\;\;\;(d+{^3\rm{He}})
\end{align*}
in the adiabatic limit. If we assume the ratio of the D to d projectile to
be 2 to 8, $U^{(c)}_e=117.5\,\rm eV$ is obtained for $E=5.01\,\rm keV$ 
as the effective screening energy. The third example is 
D($^3$He,p)$^4$He reaction at $E=4.22\,\rm keV$ , 
where the projectile is expected 
to change into a neutral He 
in about 85 $\%$ and He$^+$ in the remaining 15 $\%$ 
in the state of equilibrium charge at low energies. In the
case of $^3\rm He+\rm D$, the system converges to the ground state of
Li, while in the case of $^3\rm He^++\rm D$, the system will go into
the triplet state of Li$^+$(1s)(2s) in the probability 3/4 
and to the singlet state in the 1/4 probability \cite{bra90,mic66}. 
If the screening
energies in each state,     
\begin{align*}
203.4-(78.9+13.6)&=110.9\,[\rm eV]\;\;\;(^3{\rm He}+D)\\
137.2-(54.4+13.6)&=69.2\,[\rm eV]\;\;\;(^3\rm{He}^++D,
\mbox{singlet})\\
139.1-(54.4+13.6)&=71.1\,[\rm eV]\;\;\;(^3\rm{He}^++D,\mbox{triplet}),
\end{align*}
are averaged with the ratio, $^3\rm He+\rm D:{^3\rm He}^++\rm
D=8.5:1.5$, the effective screening energy
$U_e^{(c)}=105.3\,\rm eV$ is obtained for $E=4.22\,\rm keV$. 

TABLE \ref{tab:1} compares the effective screening energy 
$U_e^{(c)}$ which 
takes the effect of electron capture into account, 
the screening energy $U_e$ estimated by ignoring electron capture, 
and the experimental data $U_e^{exp}$\cite{dd,ali}. 
All the theoretical values were 
estimated by assuming adiabatic limit as described in the preceding 
paragraphs. 
The lower index $min$ in 
$E_{min}$ indicates 
that it is the lowest center-of-mass energy 
where experiments have so far been performed.
\begin{table}
\caption{\label{tab:1}Comparison of the screening energy estimated 
by taking electron capture by the projectile into account $U_e^{(c)}$, 
that estimated by ignoring it $U_e$ and 
the experimental value $U_e^{exp}$. }
\begin{ruledtabular}
\begin{tabular}{lccccc}   
Reaction & $E_{min}$\,(keV)  &$U_e$\,(eV) & $U_e^{(c)}$\,(eV) & 
$U_e^{exp}$\,(eV) \\
\hline
D(d,p)T & 1.62 & 22.0 & 35.7 & $25\pm5$ \cite{dd} \\
$^3$He(d,p)$^4$He & 5.01 & 119.1 &117.5 & $219\pm7$ \cite{ali}\\
D($^3$He,p)$^4$He & 4.22 & 70.6 & 105.3  &$109\pm9$ \cite{ali} \\
\end{tabular}
\end{ruledtabular}
\end{table}
The table shows that the electron capture reduces the screening energy 
for the $^3$He(d,p)$^4$He reaction. This is caused because the energy 
level of the captured electron is higher 
in the united atom than in the initial atom. 
As ref.\cite{bra90} concluded, the observed large screening energy 
in this system cannot be accounted for by the electron screening alone. 

To the contrary, the electron capture increases the screening energy 
for the D(d,p)T and D($^3$He,p)$^4$He reactions. For these reactions, 
the effective screening energy $U_e^{(c)}$, which includes the effects of 
electron capture, is as large as the experimental value $U_e^{exp}$ 
or even larger. 
In comparison, however, one should note that 
experiments are done using a molecular 
target, while our calculations were performed by approximating 
the molecule by the atom. 
Shoppa et al. \cite{sho96} showed that the screening effect strongly depends 
on the molecular orientation and that the screening energy for a  
molecular target is smaller in general than that for an atomic target 
unless a counter effect such as the reflection symmetry 
for the D(d,t)T reaction exists.  
In our case, the effective beam is the admixture of the original 
deuteron beam and the deuterium beam after electron capture. 
The screening energy in the actual molecular target will 
be larger for the deuteron beam than that estimated for the 
atomic target, while it will be smaller for the deuterium beam. 
One should modify the effective screening energy $U^{(c)}_e$ shown 
in TABLE \ref{tab:1} by taking these counter effects into account 
in order to compare with the experimental value.

\section{Effects of Coulomb explosion}

We now discuss the influence of Coulomb explosion on the
observed screening energy in 
the $^3$He(d,p)$^4$He reaction when experiments are carried out 
with diatomic ($\rm D_2^+$) or triatomic 
($\rm D_3^+$) beams instead of an atomic deuteron 
beam ($\rm D_1^+$). This is the case in \cite{ali} at ultra-low energies. 

We assume that the electron in the molecular beam is quickly lost 
in the target material and leads to the Coulomb explosion. 
We assume that the loss of electron occurs 
by the exchange process in the opposite direction of 
the electron capture by the projectile discussed in the previous section 
and assume that the electron is transferred dominantly to the state 
with a similar binding energy as that in the original molecule. 
In a simple approximation, it will then result in 
the nuclear reactions induced by the atomic deuteron 
beam with the energy $E_{\rm d}=E_{\rm 
d2}/2$ or $E_{\rm d3}/3$, 
$E_{\rm d2}$ and $E_{\rm d3}$ being the molecular beam energies in the 
laboratory system \cite{ali}. 
However, Coulomb explosion of the 
molecular beam produces an energy spread 
in the deuteron beam energy \cite{eng88}. In the case of $\rm D_2^+$ 
and $\rm D_3^+$ beams, 
the deuteron beam energy is given by  
\begin{align}
E_{\rm d}&=\frac{E_{\rm d2}}{2}+\frac{e^2}{2r}+\sqrt{\frac{e^2E_
{\rm d2}}{r}}\cos\theta~(for~ \rm D^+_2~ beam)
\label{spread2}\\
E_{\rm d}&=\frac{E_{\rm d3}}{3}+\frac{e^2}{r}+\sqrt{\frac{4e^2E_
{\rm d3}}{3r}}\cos\theta
\label{spread3}~(for~ \rm D^+_3~ beam)
\end{align} 
in the laboratory system if 
one assumes that the Coulomb energy of the molecule is 
converted into the kinetic energy of atoms. 
Here, $r$ is the internuclear distance
inside the molecule and $\theta$ is the angle between 
the incident direction of the molecule and the direction to which 
one of the deuterons is scattered in the rest frame of 
the center-of-mass of the molecule after the Coulomb explosion. 
We have ignored any intrinsic degrees of freedom of the $\rm D_2^+$ 
and $\rm D_3^+$ molecules except for $r$, which is used to estimate the 
energy release by the Coulomb explosion. 
If we denote the true screening energy at 
$E=\frac{E_{\rm d2}}{2}$ and at $E=\frac{E_{\rm d3}}{3}$ 
for the diatomic D$^+_2$ and triatomic D$^+_3$ beams, respectively, 
by $U_e^{(t)}$,  
the energy spreading given by eqs.(\ref{spread2}) and (\ref{spread3}) 
will lead to 
the following apparent screening energy $U_e^{(a)}$ 
\begin{widetext}
\begin{equation}
U_e^{(a)}=\frac{E}{\pi\eta(E)}\log\left[\frac{1}{4\pi}\int_0^\pi2\pi
\sin\theta\exp\left\{\pi\eta(E)\frac{U_e^{(t)}+(\frac{e^2}{2r}+\sqrt
{\frac{e^2E_{\rm d2}}{r}}\cos\theta)\times\frac{1.2}{2}}{E}\right\}
d\theta\right]~~~(for~ \rm D^+_2~ beam) 
\label{ua2}
\end{equation}
\end{widetext}
\begin{widetext}
\begin{equation}
U_e^{(a)}=\frac{E}{\pi\eta(E)}\log\left[\frac{1}{4\pi}\int_0^\pi2\pi
\sin\theta\exp\left\{\pi\eta(E)\frac{U_e^{(t)}+\left(\frac{e^2}{r}
+\sqrt{\frac{4e^2E_{\rm d3}}{3r}}\cos\theta\right)\times\frac{1.2}{2}}
{E}\right\}d\theta\right]~~~(for~ \rm D^+_3~ beam)
\label{ua3}
\end{equation}
\end{widetext}
where the factor $1.2/2$ transforms the energy 
from the laboratory to center-of-mass systems. 

TABLE \ref{tab:2} shows thus estimated apparent screening 
energy $U_e^{(a)}$ at each molecular beam energy. 
$E$ is the equivalent atomic deuteron energy in the
center-of-mass system at which experiments were carried out \cite{ali}.
\begin{table}
\caption{\label{tab:2}The apparent screening energy 
caused by the Coulomb explosion $U_e^{(a)}$ 
at each molecular beam energy. 
}
\begin{ruledtabular}
\begin{tabular}{cccc}
\multicolumn{2}{c}{$\rm D_2^+$}&\multicolumn{2}{c}{$\rm D_3^+$}\\
\hline
$E$(keV) & $U_e^{(a)}$(eV) & $E$(keV) & $U_e^{(a)}$(eV) \\
\hline
6.02  & 155.2& 5.01  & 213.0\\
6.45  & 154.1& 5.50  & 209.8\\
6.90  & 153.0& 6.01  & 206.7\\
7.51  & 151.8&  & \\
8.18  & 150.5&  &  \\
9.02  & 149.2&  & \\ 
\end{tabular}
\end{ruledtabular}
\end{table}
The value $U_e^{(c)}$=117.5 eV in TABLE \ref{tab:1}, 
which includes the effect of electron capture by the projectile deuteron, 
has been used for $U_e^{(t)}$, 
and $r=1.17$, 0.97 $\rm{\AA}$ for $\rm D_2^+$, $\rm D_3^+$ molecules, 
respectively \cite{eng88}. 
Interestingly, the apparent screening energy $U_e^{(a)}$=213.0 eV  
at the lowest energy $E=5.01\,\rm keV$ is 
about twice as large as the true screening energy 
and is almost the same as the experimentally reported 
value $U_e=219\pm7\,\rm eV$ 
\cite{ali}. 

\begin{figure}
\scalebox{0.8}{\includegraphics[clip]{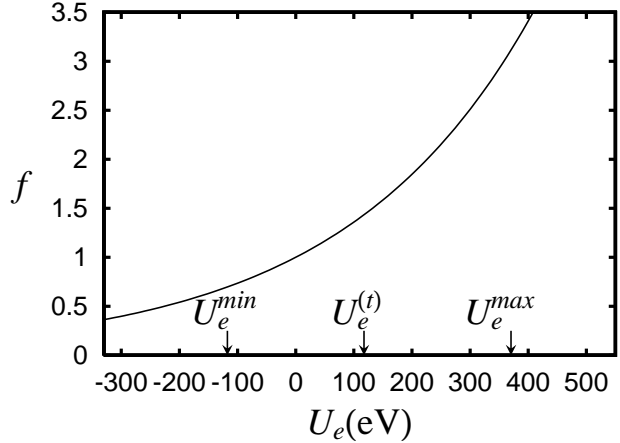}}
\caption{\label{fig:coulomb} The enhancement factor 
as a function of the screening energy. 
The positions of $U_e^{min}$, $U_e^{(t)}$, and $U_e^{max}$ are 
indicated by arrows along the abscissa. 
}
\end{figure}

The large discrepancy between the 
true and apparent screening energies can be understood 
by transforming eq.(\ref{ua3}) into 
\begin{align}
&U_e^{(a)}=\frac{E}{\pi\eta(E)}
\log \nonumber\\
&\left[\frac{1}{1.2}\int_{U_e^{min}}^{U_e^
{max}}\sqrt{\frac{3r}{4e^2E_{\rm d3}}}\exp\left(\pi\eta(E)\frac{U_e}{E}
\right)dU_e\right]
\label{ua3b}
\end{align}  
with  
\begin{equation}
U_e^{min,max}=U_e^{(t)}+\left(\frac{e^2}{r}\mp\sqrt{\frac{4e^2E_{\rm d3}}
{3r}}\right)\times\frac{1.2}{2},
\label{ueminmax}
\end{equation}
where $\mp$ corresponds to the upper index $min$ and $max$, respectively.
FIG. \ref{fig:coulomb} shows the enhancement factor of the cross 
section $f$ as a function of the screening energy, where the relevant 
region of the energy spread [$U_e^{min},U_e^{max}$] is 
indicated as well as the position of the true screening energy $U_e^{(t)}$ 
 by assuming the case of $\rm D_3^+$ beam with $E=5.01\,\rm keV$. 
The rapid increase of the enhancement factor 
within the range of the energy spread [$U_e^{min},U_e^{max}$] 
is the origin of the large apparent screening energy $U_e^{(a)}$.

Our estimate of the energy spread 
due to the Coulomb explosion 
is nearly the same as that in \cite{ali}.
The authors in \cite{ali}, however, ignored the effects of Coulomb explosion  
by attributing its justification to \cite{ms69,mv76,vm77,vm78,vs68}
which claim that the Coulomb explosion is 
much gentler and the actual energy spread is much smaller. 
\cite{mv76} reports that H$^+$-H$^+$ fragment 
pair are not observed in the dissociation of 10 keV H$^+_2$ ions 
incident on H$_2$ target.
\cite{ali} also claims that the good agreement of the data points 
obtained with the atomic and diatomic beams at nearly overlapping 
energies confirms that the effects of Coulomb explosion are negligible. 
\cite{vs68} studied the collisional dissociation of 20.4 keV H$^+_2$ in the 
H$_2$, D$_2$, He, Ne, Ar, Kr and Xe targets, that of 10.2 keV H$^+_2$ 
in the Ar target and that of 20.4 keV D$^+_2$ 
in the Ar target and has shown that the collisional dissociation of 
H$^+_2$ is dominated by the process, where it is first excited to the 
2p$\sigma_u$ state, and then dissociates and that D$^+_2$ ions behave 
in a similar way as the H$^+_2$ ions 
concerning the transitions of electrons. 
\cite{vs68} also pointed out 
the important role played by the dissociation after the excitation of 
the vibrational continuum and that by the relative orientation 
of the molecular axis to the direction of the collision. 
We have so far ignored these effects. 
It will be highly desirable to examine both experimentally and theoretically 
if the same conclusions as those in \cite{vs68} hold for the Coulomb 
explosion of $\rm D_2^+$ and $\rm D_3^+$ ions in $^{3}$He gas 
target.

\section{Summary}

We have discussed the effects of electron capture by the 
projectile ion and the Coulomb explosion in the case of molecular beams. 
We have first shown that the electron capture increases the screening energy 
for the D(d,p)T and D($^3$He,p)$^4$He reactions, and that the effective 
screening energy $U_e^{(c)}$ can get as large as the value 
observed in experiments. We have then considered the 
$^3$He(d,p)$^4$He reactions at ultra low energies induced by 
$\rm D^+_2$ and $\rm D^+_3$ beams, and have shown that the Coulomb explosion 
can lead to a large apparent screening energy, which is almost twice 
as large as the adiabatic limit and matches with the value reported 
by the experimental analysis which ignores the effects of Coulomb explosion. 
We have assumed the prompt explosion of the molecular ions by the loss 
of electrons due to exchange process.
In view of former studies \cite{ms69,mv76,vm77,vm78,vs68}, however,  
more detailed studies of the Coulomb explosion will be 
needed in order to draw a definite conclusion.
As we have implicitly postulated, 
we are especially interested in the Coulomb dissociation caused by the 
transfer of electrons from the molecular projectiles 
to the target gas in the same mechanism 
as that to establish the equilibrium charge and also  
the neutralization of thus synthesized ions.  
Whether these processes occur significantly or not when the molecular 
$\rm D^+_2$ and $\rm D^+_3$ beams propagate through the $^3$He gas target 
will be discussed in a subsequent paper. 

\begin{acknowledgments}
We thank Prof. T. Itahashi and Dr. S. Kimura 
for useful discussions.
\end{acknowledgments}


\begin{thebibliography}{99}

\bibitem{alr87}  H. J. Assenbaum and K. Langanke and C.
Rolfs, Z. Phys. A{\bf 327}, 461 (1987).

\bibitem{rs95}  C. Rolfs and E. Somorjai, Nucl. Inst. Meth. Phys. 
Res. B{\bf 99} 297 (1995).

\bibitem{ktab03} S. Kimura, N. Takigawa, M. Abe and D.M. Brink, 
Phys. Rev. C67 022801(R)-1-5 (2003).
 
\bibitem{gs-e}  R. Golser and D. Semrad, Phys. Rev. Lett. {\bf 66},
  1831 (1991).

\bibitem{gs-t}  P.L. Grande and G. Schiwietz, Phys. Rev. A{\bf 47},
1119 (1993).

\bibitem{bp}  C. A. Bertulani and D. T. de Paula, Phys. Rev. C{\bf
  62}, 045802 (2000).

\bibitem{barker}  F. C. Barker, Nucl. Phys. A{\bf 707}, 277 (2002).

\bibitem{junker}  M. Junker, {\it et al.} Phys. Rev. C{\bf 57}, 2700 (1998).

\bibitem{bra90}
L. Bracci et al., Nucl. Phys. A{\bf 513}, 316 (1990).


\bibitem{eng88}
S. Engstler et al., Phys. Letts. B{\bf 202}, 179 (1988).


\bibitem{pdg} Particle Data Group, Particle Physics Booklet.

\bibitem{kato04} Y. Kato, Master thesis, Tohoku University (2004). 

\bibitem{all58}
S. K. Allison, Rev. Mod. Phys. {\bf 30}, 1137 (1958).

\bibitem{wei}
M. Weissbluth, {\it Atoms and Molecules} (Academic Press, New York,
1980).
\bibitem{com}
S. E. Koonin and D. C. Meredith, {\it Computational Physics}
(Addison-Wesley, Redwood City, CA, 1990).

\bibitem{mic66}
H. H. Michels, J. Chem. Phys. {\bf 44}, 3834 (1966).

\bibitem{dd}
U. Greife et al., Z. Phys. A{\bf 351}, 107 (1995).

\bibitem{ali}
M. Aliotta et al., Nucl. Phys. A{\bf 690}, 790 (2001).


\bibitem{sho96}  T. D. Shoppa, M. Jeng, S. E. Koonin, K. Langanke and R.
Seki, Nucl. Phys. A{\bf 605}, 387 (1996).


\bibitem{ms69} B. Meierjohann and W. Seibt, Z.Phys. {\bf 225}, 9 (1969).

\bibitem{mv76} B. Meierjohann and M. Vogler, J. Phys. B{\bf 9}, 1801 (1976).

\bibitem{vm77} M. Vogler and B. Meierjohann, Z. Phys. A{\bf 283}, 11 (1977).

\bibitem{vm78} M. Vogler and B. Meierjohann, J. Chem. Phys. {\bf 69}, 2450 
(1978).

\bibitem{vs68} M. Vogler and W. Seibt, Z. Phys. {\bf 210} 337 (1968).


\end{thebibliography}
\end{document}